\documentstyle[12pt]{article}

\newcommand{\ie}{{\it i.e.}}

\newcommand{\ccbar}{c\bar{c}}
\newcommand{\QQbar}{Q\bar{Q}}
\newcommand{\qqbar}{q\bar{q}}
\newcommand{\jp}{J/\psi}

\newcommand{\PLB}[3]{\mbox{}Phys. Lett. {\bf B{#1}}, {#2} ({#3})}
\newcommand{\NPB}[3]{\mbox{}Nucl. Phys. {\bf B{#1}}, {#2} ({#3})}
\newcommand{\PRL}[3]{\mbox{}Phys. Rev. Lett. {\bf {#1}}, {#2} ({#3})}
\newcommand{\PRD}[3]{\mbox{}Phys. Rev. {\bf D{#1}}, {#2} ({#3})}

\newcommand{\etal}{{\em et al.}}

\begin{document}

\thispagestyle{empty}
\begin{flushright}
   \vbox{\baselineskip 12.5pt plus 1pt minus 1pt
         SLAC-PUB-6577 \\
         July 1994 \\
         (T/E)
             }
\end{flushright}

\begin{center}
{\bf Anomalous Charm Production at Large $x_{F}$
\footnote{Invited talk presented at Workshop on the Future of High
Sensitivity Charm Experiments: Charm2000, Fermilab, Batavia, Il.,
June 7-9, 1994}}

\vskip 1\baselineskip

W.-K. Tang\footnote{Work supported by the Department
 of Energy, contract DE-AC03-76SF00515} \\

{\normalsize \em
  Stanford Linear Accelerator Center} \\ {\normalsize \em Stanford
  University, Stanford, CA 94309}
\end{center}

\medskip

\begin{abstract}
\noindent
We show that the new QCD production mechanisms which were proposed
by S. J. Brodsky, P. Hoyer, A. H. Mueller and the author can explain at
least
some of the anomalous behavior of open and/or closed charm
production at large $x_{F}$.
\end{abstract}

\section{Introduction}
Charm production at large
$x_{F}$ is a very fascinating regime which provides a lot of information
about the internal structure, especially the higher Fock state
components of the projectile in the question \cite{BHMT,hoyer}. No matter
whether it is deep
inelastic scattering, pion nucleus collisions, open charm or hidden
charm production \cite{EMC,CIP,E537,NA3,WA82,E769,NA3a,E789,E772,HVS}, all of
them show
anomalous behavior which cannot be explained
by leading twist PQCD. In fact, in the $x \rightarrow 1$ limit, there
is a new hard scale $\Lambda_{QCD}^2/(1-x)$, and the corrections to
leading twist terms are of order $\Lambda_{QCD}^2/(1-x)Q^2$. Actually,
in the
combined limit,
\begin{equation}
\left.
\begin{array}{l}
   Q^2\rightarrow \infty \\
   x \rightarrow 1
   \end{array}
\right\} \;\;\;\;\;\; \mbox{with $\;\mu^2 \equiv (1-x)Q^2 \;$fixed,}
\label{combined}
\end{equation}
the twist expansion breaks down and higher twist terms are no longer
suppressed and can become dominant.

This paper is organized as follow: in section 2, we will review some of
the experimental data which
shows that  higher
twist effects are important at large $x_{F}$. It is well known that higher
twist
terms are
suppressed by ${\cal O} (1/M^2)$ so it raises the question why they become
dominant at large $x_{F}$. Thus, we need to understand the physical
origin of the suppression of higher twist terms at moderate $x_{F}$
and this is reviewed in section 3. We then explain
in section 4
why
in the new limit, $x_{F} \rightarrow 1$ and $M^{2} \rightarrow \infty$
but with
$\mu^2 = (1-x_{F}) M^2$ fixed, higher twist terms are not
suppressed. We will argue in section 5 that in this new QCD
limit, the cross section for freeing the $\ccbar$ pair is not small
due to the fact that
 the dominant contribution comes from peripheral processes in
which  slow  spectator quarks interact with the target. These
new production mechanisms are then applied to different processes and can
explain at least some of anomalous behavior observed. Finally, we give
our conclusion in the last section.

\section{Anomalous behavior of open and/or hidden charm production at
large $x_F$}

{\rm \bf (a)} \hspace{0.7cm}
 It is reported by EMC \cite{EMC} that the  $c(x)$ distribution measured at
large
$x_{bj}$ is
 anomalously high.  The CERN measurements disagree with photon-gluon fusion
by a factor of 20 to 30 at $Q^2 = 75$ GeV$^2$ and $x_{bj} = 0.422$
as shown in Fig. 1a and b. One should  notice that the measured
$x_{bj}$ is the fractional longitudinal momentum for one charm quark
only. The total $x_{bj}$ for the $\ccbar$ pair should be nearly double
which is $0.85$, very close to $1$. In Fig. 1a and b, one can see that
photon-gluon fusion fits the data well for $x_{jb}$ less than
$0.3$, but badly for $x_{bj}$ larger than $0.4$ .

\begin{figure}[htb]
\vspace{4.5in}
\caption {Fig. 1a shows the  value of $F^{\ccbar}_2$ versus $Q^2$ for
fixed $x$: $\circ$ (x=0.00422), $\times (x=0.00750)$, $\bigtriangleup
(x=0.0133)$, $\otimes (x=0.0237)$, $\diamond (x=0.0422)$, $\bullet
(x=0.0750)$, $\Box (x=0.133)$, $\oplus (x=0.237)$, $\bigtriangledown
(x=0.422)$. Please notice the large disagreement between experimental data
$\bigtriangledown$
and the PGF prediction (smooth curves). In Fig. 1b the charm quark momentum
density
distribution $xc(x)$ as a function of $x$ and $Q^2$ is plotted. The PGF
prediction (smooth curves) does not fit the data when $x$ is large. The
data is from Ref.\protect\cite{EMC}.}
\label{first-figure}
\end{figure}

\noindent {\rm \bf (b)} \hspace{0.7cm}
A sudden change in polarization of the $J/\psi$
is reported by CIP \cite{CIP} and E537 \cite{E537} at large $x_F$ in $\pi N$
collisions.
The polarization of the $\jp$ is determined by the angular distribution of its
decay muons in the $\jp$ rest frame. By rotational symmetry and parity, the
angular distribution of massless muons, integrated over the azimuthal angle,
has
the form
\begin{equation} \frac{d\sigma}{d\cos\theta} \propto 1 + \lambda \cos^2
\theta \label{lambda}
\end{equation}
where  $\theta$ is the angle between the $\mu^+$ and the projectile
direction (\ie, in the Gottfried--Jackson frame). The parameter
$\lambda$ is directly related to the polarization of the $\jp$ particle,
\ie,
\begin{equation}
\lambda = \left\{ \begin{array}{ll}
                  1  & \;\;\;\;\;\;\; \mbox{transverse $\jp$} \\
                  0  & \;\;\;\;\;\;\; \mbox{unpolarized $\jp$} \\
                  -1 & \;\;\;\;\;\;\; \mbox{longitudinal $\jp$.}
                  \end{array}
          \right.
\end{equation}
In Fig. 2, where $\lambda$ is plotted against $x_{F},$ one can clearly
see that the polarization of the produced $\jp$ changes sharply from
unpolarized to longitudinally polarized around $x_{F} \sim 0.85$. This
dramatic effect is inconsistent with leading order QCD.

\begin{figure}[htb]
\vspace{3.8in}
\caption {CIP data: $x_{F}$ dependence of $\lambda$ fitted to the $\jp$
decay. Please notice the sudden change of the $\jp$ polarization
around $x_{F} \sim 0.85$.}
\label{second-figure}
\end{figure}

\noindent {\rm \bf (c)} \hspace{0.7cm}
The measurement from NA3 \cite{NA3} shows that double $J/\psi$ pairs are
hadroproduced only at large $x_F.$ In the NA3 experiment, 6 $\psi\psi$
are found at 150 GeV and 7$\psi\psi$ at 280 GeV in  $\pi^-$
beam scattering with a platinum target. In table 1, we list the $x_{F}$
of the $\psi\psi$ pair of all 13 events in ascending order. The mean
$x_{F}$ of the pair is $0.71$ (150 GeV) and $0.53$ (280 GeV) which is
very large.

 \begin{table}[htb]
 \begin{center}
\begin{tabular}{|c|ccccccc|}     \hline
 $P_{\pi}$ & \multicolumn{7}{c|}{$x_{F}$ of $\psi\psi$ pair} \\  \hline
  150 GeV & 0.58 & 0.61 & 0.75 & 0.75 & 0.77 & 0.78 &  \\ \hline
  280 GeV & 0.39 & 0.47 & 0.47 & 0.48 & 0.51 & 0.65 & 0.75  \\ \hline
\end{tabular}
\end{center}
\caption{The $x_{F}$
of the $\psi\psi$ pair of all 13 events in ascending order.
The data are from Ref.
\protect\cite{NA3}.}
\end{table}

 The data also indicates strong
correlations in the production mechanisms. The transverse
momentum of the $\psi\psi$ pair  is $0.9 \pm 0.1$ GeV for the $280$ GeV beam;
whereas uncorrelated pairs from a Monte Carlo study have a much larger mean
transverse momentum of $1.7$ GeV. Also amazingly, the mean value of the
individual $\jp$ transverse momenta in the $\psi\psi$ events, $1.5$ GeV, is
significantly higher than the mean transverse momentum of the $\psi\psi$
pair, so there is strong correlation of between the transverse momenta
of two $\jp$'s produced.

To make a quantitative statement about the correlation, we should
compared the measured
double $\jp$ cross section per nucleon $\sigma_{\psi\psi}$
with $A^{1/3}(\sigma_{\psi}/\sigma_{tot})^2 \sigma_{tot}$, which is the
theoretical
estimate assuming that the $\psi$'s are produced uncorrelated. Here
$\sigma_{\psi}/\sigma_{tot}$ is the probability of producing a $\jp$ in
a
pion nucleon collision. The extra $A^{1/3}$ dependence takes into
account the nuclear effect. In table 2, we compare the two cross
sections and find that the theoretical prediction is off by three
orders of magnitude! This strongly indicates that we need a new production
mechanism in order to account for the large disagreement. The leading
charm hadroproduction and the nuclear dependence, being reviewed
in the following paragraphs, give us hint of the nature of this new
mechanism.

\begin{table}[htb]
\begin{center}
\begin{tabular}{|c|c|c|c|c|}
  \hline
$P_{\pi}$  & $\sigma_{\psi\psi}$ [pb] & $\sigma_{\psi}$ [nb]
    & $\sigma_{tot}$ [mb] &
    $A^{1/3}(\sigma_{\psi}/\sigma_{tot})^2 \sigma_{tot}$ [$10^{-2}$pb]\\ \hline
  150 GeV & $18\pm 8$  & 6.5  & $\sim 25$ & 1.0 \\ \hline
  280 GeV & $30\pm 10$ & 8.7 & $\sim 25$ & 1.7 \\ \hline
\end{tabular}
\end{center}
\caption{ Cross sections per nucleon for double $\jp$ production
in $\pi^- N$ collisions and the
theoretical prediction assuming the $\jp$'s are produced uncorrelated.
The data is from Ref.
\protect\cite{NA3}.}
\end{table}

\noindent {\rm \bf (d)} \hspace{0.7cm}
Dramatic leading particle effects in hadronic $D$
production are observed by WA82
\cite{WA82} and E769 \cite{E769} experiments.
In $\pi^- (\bar{u}d)$ interactions with hadrons or nuclei, the
$D^-(\bar{c}d)$
and $D^0 (c\bar{u})$ are referred to as ``leading" charm mesons while
the $D^+(c\bar{d})$ and $\overline{D}^0 (u\bar{c})$ are ``nonleading". The
asymmetry between leading and nonleading charm, which has been used in
the analyses of the WA82 and E769 collaborations, is defined as
\begin{equation}
{\cal A} = \frac{\sigma(\mbox{leading}) -\sigma(\mbox{nonleading})}
           {\sigma(\mbox{leading}) +\sigma(\mbox{nonleading})}
\end{equation}
Both experiment find that the measured ${\cal A}(x_{F})$ increases from
$\sim 0$  for $x_{F} \leq 0.4$  to $\sim 0.5$ around $x_{F} = 0.65$
(Fig. 3). Therefore, the leading charm asymmetry is localized at large
$x_{F}$ only.

\begin{figure}[htb]
\vspace{3.2in}
\caption {Leading charm asymmetry versus $x_{F}$. A substantial asymmetry
is observed at large $x_{F}$.}
\label{third-figure}
\end{figure}

According to leading twist QCD, the  hadroproduction cross section of $D$
mesons is given by
\begin{equation}
\frac{d\sigma_{AB\rightarrow DX}}{dx_{a}dx_{b}dz_{1}} \propto \sum_{ab}
  f_{a/A}(x_{a})  f_{b/B}(x_{b}) \hat{\sigma}_{ab\rightarrow \ccbar}
  D_{D/c}(z_1).
\label{asymmetry}
\end{equation}
The structure functions of the initial hadrons, $ f_{a/A}(x_{a})$, are
process independent while the fragmentation functions $ D_{D/c}(z_1)$
are independent of the quantum numbers of both the projectile and the
target. Thus, leading twist QCD predicts the  leading charm
asymmetry to be nearly zero.\footnote{Next-to-leading order calculation do give
rise to a
small charge asymmetry between $\bar{c}$ and $c$ production due to
$qg$ and $q\bar{q}$ interference \cite{Beenakker,NDE}.} The observed
large leading charm asymmetry
breaks QCD factorization which strongly suggests that it is a higher twist
effect.

\noindent {\rm \bf (e)} \hspace{0.7cm}
The data \cite{NA3a,E772,E789} on $\jp$ production in hadron-nucleus
collisions exhibits a surprising result. The NA3 and E772
 data give direct evident
for the breakdown of the leading twist approximation at large $x_{F}$.
Following the argument of Ref.\cite{HVS},  by the factorization theorem,
the cross section of  $\jp$
production in $\pi A$ collisions is,

\begin{equation}
\frac{d\sigma_{\pi A\rightarrow \jp X}}{dx_{1}dx_{2}} =
  f_{a/\pi}(x_{1})  f_{b/A}(x_{2}) \hat{\sigma}({ab\rightarrow \jp})
\label{factor}
\end{equation}

For simplicity we just assume gluon gluon fusion to be dominant. For
$\sqrt{s} \gg M_{\psi}^2$ and $x_{F} > 0$, approximately $x_1 \simeq
x_F$, $x_2 \simeq M^{2}_{\psi}/x_F s$.  In the factorized formula
(\ref{factor}), the nuclear $A$-dependence appears only through the
target function $ f_{b/A}(x_{2})$. Hence, ratios  $R = A\sigma(pp\rightarrow
 \jp +X)/\sigma(pA\rightarrow \jp
+X)$ of $\jp$ production  should be independent of c.m. energies $\sqrt{s}$
when $\sqrt{s}$ and $x_F$ varied in such a way as to keep $x_2$
constant. However, as
shown in Fig. 4, the  NA3 data \cite{NA3a} shows that the ratio for H/Pt is
consistent with {\it Feynman scaling}, \ie, scales with $x_F$ but not
with $x_2$. A clear energy dependence is seen at  small values of
 $x_2$. Thus the leading twist factorization fails at large Feynman $x$
 of the $\jp$, since $x_2 \simeq M^{2}_{\psi}/x_F s$. A similar
result was observed by combining $pA$ data from NA3 and
E772 \cite{E772}.

\begin{figure}[htb]
\vspace{4.3in}
\caption {The ratio $R = A\sigma(pp\rightarrow \jp +X)/\sigma(pp\rightarrow \jp
+X)$ of inclusive $\jp$ production on Hydrogen and Platinum \protect\cite
{NA3a}.
In (a) the ratio is plotted as a function of $x_F$ of $\jp$, and in (b)
as a function of $x_2$.}
\label{fourth-figure}
\end{figure}

The same anomalous behavior is also observed if one studies the nuclear
$A$-dependence of the $\jp$ production cross section through the
parametrization  $\sigma_A = \sigma_p A^{\alpha}$. The effective power $\alpha$
at
different energies show that indeed $\alpha = \alpha(x_F)$, \ie, the
nuclear suppression obeys Feynman scaling \cite{E772}, and is not a
function of $x_2$. The power $\alpha$ decreases from 0.97 at $x_F = 0$
to 0.7 as $x_{F} \rightarrow 1$ \cite{NA3a,E772,E789} (see Fig. 5),
\ie, becomes surface dominated at large $x_F$.

\begin{figure}[htb]
\vspace{3.5in}
\caption {The effective power $\alpha$ of the $A-$dependence of $\jp$
production : E789 ($\bullet$) and E772 $(\circ)$ data.}
\label{fifth-figure}
\end{figure}

The above nuclear $A$-dependence and the leading charm asymmetry
 directly contradict leading-twist PQCD factorization and
suggest that higher twist effects play an important role at large $x_F$.
But it is a well known fact that higher twist effects are suppressed by
${\cal O} (1/M^2)$. This raises the question of how the higher twist
effects survive and become dominant at large $x_F$. In the next
section we will review the physics of higher twist terms and point out
a way to overcome the usual suppression.

\section{Physical Picture of Higher Twist Terms}

Let us take the well known process, Deep Inelastic Scattering (DIS), to
illustrate the physics of `leading twist' and `higher twist' terms.
In leading twist diagrams (Fig. 6a), only the active (hit) parton interacts
 with
the external photon and there is no connection between the spectator
partons and the active parton. On the other hand, there are strong
interactions between the active parton and the spectator partons in
higher twist diagrams (Fig. 6b).

\begin{figure}[htb]
\vspace{2.5in}
\caption {DIS scattering: (a) leading twist and (b) higher twist.}
\label{sixth-figure}
\end{figure}

If we take the `infinitive momentum frame' in which the parton language
 is valid, the proton is boosted to  very high momentum along the $z$
axis with four momentum given by $P = (p + m^2/2p, \underline{0}, p)$.
In this frame, the photon momentum can be taken as $q =(Q^2 p/m^2
x,\underline{q},
Q^2 p/m^2 x)$ and the virtuality of the photon is $Q^2=
\underline{q}^2$; \ie, the resolving power in transverse
dimension. In other words, the transverse dimension of the partons that
interact, directly or indirectly, with the photon is of the order of
$1/Q$. With the above pictures in mind, it is easy to understand why
the higher twist terms are suppressed by $1/Q^2$ in the usual Bjorken
limit $Q^2\rightarrow \infty$ with $x$ fixed. As the interaction time
$\tau$ of
the hard subprocess $eq \rightarrow eq$
scattering is very short, only of the order of $1/Q$, any interaction between
the
active parton and the spectator partons must occurs within this short
time interval $\tau$ and so they must be within transverse
distance of $r_{\perp} \sim 1/Q$ (Fig. 7). This immediately leads to
the conclusion that higher twist terms
are suppressed by $1/Q^2 $ as
the probability of finding
two partons with dimension $1/Q^2$ within an area of $1/Q^2$ is given by the
geometrical factor
$1/Q^2 R^2$, with $R$ the size of proton.

\begin{figure}[htb]
\vspace{2.0in}
\caption {The transverse view of the partons in the hadron in the
infinite momentum frame.}
\label{seventh-figure}
\end{figure}

However, there is an exception to the above conclusion. Suppression of
higher twist terms depends a lot on the size of proton, which is of
order of $1$ fm, much larger than the size of the parton, which is $1/Q$.
If somehow the
initial proton or meson is already very small, of the same size as the
parton, then there is no suppression. But how can that be realized? The
answer to that question lies on the large $x$ kinematic region
and we will review that region in the next section.

\section{The combined limit : $x \rightarrow 1$ and $Q^2 \rightarrow
\infty$ with  $(1-x)Q^2$ fixed}

In the large $x$ kinematic region, besides the usual hard scale $Q^2$,
another new scale $\Lambda^2_{QCD}/(1-x)$, which reflects the hardness of
this new limit, emerges \cite{hoyer,BHMT}. In fact, this new hard scale
actually
is the
transverse size of the meson; \ie, $r^{2}_{\perp} \sim (1-x) /
\Lambda^2_{QCD}$. If the two scales are comparable; \ie, taking
the combined limit as in equation (\ref{combined}),
higher twist contributions will not be suppressed assuming $\mu^2 \sim
\Lambda^2_{QCD}$.   In this new limit,  higher and leading twist are of
the same order,
\[ \frac{1/Q^2}{r^{2}_{\perp}} \sim \frac{\Lambda^2_{QCD}}{(1-x)Q^2} \sim
\frac{\Lambda^2_{QCD}}{\mu^2} \sim 1. \]

But how does the new hard scale  $\Lambda^2_{QCD}/(1-x)$ emerge in the
limit  $x \rightarrow 1$? Let us consider Fig. 8 which gives the probability
amplitude for the $x \rightarrow 1$ perturbative distribution of the meson. The
soft
non-perturbative distribution is described by the wavefunction
$\phi(yp,n_{\perp})$ which is suppressed in the extreme kinematic limit
$y \rightarrow 0, 1$ or $n_{\perp} \rightarrow \infty$. The
perturbative contribution comes from  diagrams where one or more
gluons are exchanged between the two quarks. For simplicity,
we just consider exchanging one gluon between the quarks in Fig.~8.

\begin{figure}[htb]
\vspace{2.0in}
\caption {The $x\rightarrow 1$ limit of a hadron structure function
generated by perturbative gluon exchange.}
\label{eighth-figure}
\end{figure}

The separation between the two quarks $r_{\perp}$ can be estimated by
considering the intermediate state $\qqbar g$. The virtuality of this
state is given by
\begin{eqnarray}
2p \Delta E_{\qqbar g} & \simeq& -\frac{m_{q}^2+n_{\perp}^2}{y} -
                       \frac{m_{q}^2+k_{\perp}^2}{1-x} -
                       \frac{(n_{\perp}-k_{\perp})^2}{x-y} \nonumber \\
                       & \simeq & \frac{m_{q}^2+k_{\perp}^2}{1-x}
\end{eqnarray}
when
\begin{equation}
 n_{\perp}^2 \leq {\cal O} (\frac{k_{\perp}^2}{1-x}).
 \label{eq:hard}
\end{equation}
Since $\Delta E_{\qqbar g}$ is independent of $n_{\perp}$,  the
perturbative tail is
\begin{eqnarray}
\int^{\frac{k_{\perp}^2}{1-x}} dn_{\perp} \phi(yp,n_{\perp})&\simeq&
   \int^{\infty} dn_{\perp} \phi(yp,n_{\perp}) \nonumber \\
   &\simeq& \phi(yp,r_{\perp} = 0)
\end{eqnarray}
which shows that the transverse distance between the two quarks is
$r_{\perp}^2\sim (1-x)/k_{\perp}^2$, very compact at the moment of
creation. For a typical value, $k_{\perp} \sim \Lambda_{QCD}$, we find the new
hard scale  $\Lambda^2_{QCD}/(1-x)$ as promised.

Another interesting physical quantity is the transverse distance
$R_{\perp}$
between
the two quarks after the exchange of the gluon, \ie, when one of the
quark carries nearly all the longitudinal momentum. The life time of
this intermediate state is very brief,
\begin{equation}
\Delta \tau \simeq \frac{1}{\Delta E_{\qqbar}}
            \simeq \frac{2p(1-x)}{k_{\perp}^2+m_{q}^2}
\end{equation}
Nevertheless, during this short life time, the `slow' quark can move a
transverse distance
\begin{equation}
R_{\perp} \simeq v_{\perp}\Delta \tau =
\frac{k_{\perp}}{p(1-x)}\frac{2p(1-x)}{k_{\perp}^2+m_{q}^2} \simeq
\frac{2k_{\perp}}{k_{\perp}^2+m_{q}^2}
\end{equation}
which for $k_{\perp} = {\cal O}(\Lambda_{QCD})$ can be of the order of
1 fm. Hence, the specific large $x$ kinematic region selects a very compact
Fock
state component of the meson at the moment of creation and then it expands very
quickly to its normal size of 1 fm. The large transverse size
$R_{\perp}$ of the light quarks has a very important implication in the
production of heavy quarks.

\section{Dynamics in the new QCD Limit}

In the previous section, we showed that there is a new scale
$\Lambda_{QCD}^2/(1-x)$ at large $x$ and that the transverse size of the
light quarks $R_{\perp}$ can be as large as 1 fm. In this section, we
want to exploit these properties in the production of heavy quarks at
large $x$. As the transverse size $R_{\perp}$ of the light quarks
is very large, one can imagine that the heavy quark pair
can be freed easily by deflecting the slow  light
quark. This phenomenon has been studied in Ref.\cite{BHMT} in the case
of heavy quark production on nuclear target.
  The new limit in this case is defined by:

\begin{equation}
\left.
\begin{array}{l}
   M^2\rightarrow \infty \\
   x \rightarrow 1
   \end{array}
\right\} \;\;\;\;\;\; \mbox{with $\;\mu^2 \equiv (1-x)M^2 \;$fixed}
\label{new}
\end{equation}
where $M$ is the mass of the heavy quark pair. To understand the physics in
this
new limit, let us consider the
 ``extrinsic" and ``intrinsic" diagrams as shown in Figs
9a and 9b. In the extrinsic diagram the produced heavy quark pair
couples directly to only one parton in the projectile while in the
intrinsic case it couples to several.

\begin{figure}[htb]
\vspace{4.5in}
\caption {Leading order diagrams in heavy quark production in the new limit
(\protect\ref{new}): (a)
extrinsic diagram and (b) intrinsic diagram.}
\label{ninth-figure}
\end{figure}

The energy difference in the
extrinsic diagram is given by

\begin{equation}
2p \Delta E \sim \frac{k_{\perp}^2}{1-x} + M^2.
\label{energydifference}
\end{equation}
The first term $k_{\perp}^2/(1-x)$ comes from the effectively stopped
light valence quarks $\qqbar$ as the produced $\QQbar$ pair carries
almost all of the momentum ($x \rightarrow 1$) while the second term
comes from the virtuality of the gluon which is of the order of the mass of
the heavy quark pair.  In
order to get a large production cross section, the energy difference should be
minimized
and thus the two terms in equation (\ref{energydifference}) are of the
same order, \ie,

\begin{equation}
k_{\perp}^2 \sim (1-x)M^2 = \mu^2
\end{equation}

Now we have a very nice result. The transverse momentum square of the light
quarks are of the order of $\mu^2$ and so these states can be resolved
by a target gluon of transverse momentum $l_{\perp}$ of  order of
$\mu$. Hence the hardness of the scattering  from the target
is not $M^2$ as one would expect in the leading twist calculation,  but
instead it is $\mu^2=(1-x)M^2$. Actually, the transverse size of the
stopped
light quark pair is given by $1/k_{\perp} \sim 1/\mu$. This explains
why the scattering dominantly occurs off the light quarks. Therefore,
we can conclude that heavy quarks can be, and are, produced at large
$x$ by soft peripheral scattering and so the cross section is large.
These new production mechanisms can help to explain the various
anomalous behaviors of charm production as described in section 2.

In leading twist diagrams, the usual lowest order diagram describing
the fusion process $gg \rightarrow \QQbar$ is shown in Fig. 10.
Although the size of the light quark pair is large, the heavy quark
pair $\QQbar$ still has a small transverse size $h_{\perp} \sim 1/M$.
A target gluon can resolve the $\QQbar$ pair only provided that it has
a commensurate wavelength, \ie, $l_{\perp} \sim M$ as indicated in
Fig.~11. This is much large than the $l_{\perp} \sim \mu$ required to
resolve the light quarks. Hence the leading twist is actually
suppressed by $1/M$ compared to the new mechanisms in the new limit.

\begin{figure}[htb]
\vspace{2.2in}
\caption {Leading twist diagram in heavy quark production: $gg
\rightarrow \QQbar$.}
\label{tenth-figure}
\end{figure}

One can also go through the same argument as described in the
previous section and conclude that the Fock state of the projectile
hadron from which the heavy pair is produced has a small transverse size
$r_{\perp}^2 \sim (1-x)/\mu^2 \sim 1/M^2$. Because of the smallness
of the transverse size,  the intrinsic diagram as shown in Fig.~9b,
where an extra gluon is attached to the heavy quark pair, is not
suppressed relative to the extrinsic diagram Fig.~9a.  Therefore, the
distinction between extrinsic and
intrinsic processes essentially disappears.

\section{Applications of the New Mechanisms}
Let us summarize the  physics in the new limit (\ref{new})
before we go on to apply it to the anomalous charm production. In the
combined limit, the Fock states are very compact and small. The
transverse radius square of the states has a typical value of
$(1-x)M^2$. Because of the
compactness of the Fock states, intrinsic diagrams and extrinsic
diagrams are of the same order. But the intrinsic diagrams can
numerically dominate the extrinsic contributions because of the large
combinatic factor.  The heavy quark pair can be freed easily by
stripping away the slow light spectator quark in the projectile through
an interaction with the target. The hardness
scale of the collision is given by $(1-x)M^2$. It is a soft peripheral
process and so the cross section is large. This picture can
provide a QCD framework for understanding the puzzling features of the
large $x$ data mentioned in section 2:

\noindent {\rm \bf (a)} \hspace{0.7cm}
The larger than expected charm structure function of the nucleon at
large $x_{bj}$
reported by EMC \cite{EMC} can be understood by the large intrinsic charm
contribution in the proton.  In this case, $Q^2 = 75$ GeV$^2$ (for the
data point with $x_{bj} = 0.422$) is
fixed and the photon can
resolve the charm quark easily. But as discussed above, the intrinsic
production of the charm pair at large $x_{F}$ (which is nearly twice of
$x_{bj}$)  can
numerically dominate the usual extrinsic production considered in PGF
calculation and boost up the charm structure function a lot.

\noindent {\rm \bf (b)} \hspace{0.7cm}
The longitudinally polarized $\jp$ at large $x_F$ in $\pi N$ collisions
has a natural explanation by the new production mechanisms \cite{VHBT}.
The dominant contributions  to the polarization of $\jp$ at large
$x_{F}$ are the intrinsic diagrams as shown
in Fig.~11. The initial state pion valence quarks naturally have
opposite helicities. There is a factor $1/(1-x_F)$ enhancement for the
emission of transversely polarized gluon with same sign of helicity
as the radiated quark. Thus the two gluons coupled to the charm pair
have opposite helicity. In order to form a bound state, the transverse momenta
of
the gluons and thus of the charm pair should be small. In that case, the
initial $J_z =0$ and thus the formed $\jp$ is in longitudinally polarized
state. Here, we have made the assumption that the formation of the $\jp$
through the radiation of an extra gluon does not change the
polarization of the charm quarks. In the large $x_{F}$ limit, the
radiated gluon must be soft and this justifies our assumption.

\begin{figure}[htb]
\vspace{2.4in}
\caption {Dominant diagram in heavy quark production. The plus and
minus signs refer to the particle helicities.}
\label{eleventh-figure}
\end{figure}

The
counting rules in powers of $1-x_{F}$ are presented in Ref.\cite{VHBT}.
The cross section for producing a longitudinally polarized (Fig.~12a)
and a transversely polarized (Fig.~12b) $\jp$
is proportional to $(1-x_F)^3$ and $(1-x_F)^4$ respectively. We find that
the basic reason for the dominance of the intrinsic
polarization amplitude (Fig. 12a) is that it allows two helicity flips
of the heavy quarks, each contributing a power of $M \sim
1/\sqrt{1-x_F}$ in our analysis. Thus, the longitudinal polarization of the
$\jp$ at large $x_F$ is mainly due to intrinsic charm production
mechanisms.

\begin{figure}[htb]
\vspace{2.0in}
\caption {The production of (a) longitudinal and (b) transverse $\jp$.}
\label{twelveth-figure}
\end{figure}

\noindent {\rm \bf (c)} \hspace{0.7cm}
The anomalous double $\jp$ production can be understood qualitatively by
considering
intrinsic production as shown in Fig.~13. The two $\jp$'s produced as shown
in the diagram clearly are
strongly correlated. The cross section for freeing the pairs becomes large at
large
$x_F$ as it is a soft peripheral scattering from the target. This  helps to
explain
 why the $\jp$ pairs produced at large $x_F$
only. Intrinsic charm production (Fig.~13) has  another nice feature. The total
transverse momentum
square of the  $\jp$ pair
is of the order of $\mu^2=(1-x_F)M^2$ only, \ie, of the same
order as that of the light quarks. However, the individual $\jp$'s can
have
transverse momenta up to the mass scale $M$. In fact, if one uses the
measured mean value of $x_F$, which is 0.53, from the NA3 280 GeV beam
data and
the measured mean transverse momentum of the individual $\jp$'s,
$M \sim k_{\perp} = 1.5$ GeV, one gets $\mu = 1.0$ GeV which is close
to
the measured value of $<p_{\perp}^{\psi\psi}>= 0.9\pm 0.1$ GeV. Obviously, one
cannot take
this
number too seriously. Nevertheless, it  indicates that all these
features fit
nicely with the data and the proposed new mechanisms may play
an important role in double $\jp$ production.

\begin{figure}[htb]
\vspace{2.2in}
\caption {Intrinsic diagram that may account for
  double $\jp$ production at large $x_{F}$.}
\label{thirteenth-figure}
\end{figure}

\noindent {\rm \bf (d)} \hspace{0.7cm}
The leading charm asymmetry has been studied in detail by R. Vogt and S. J.
Brodsky
\cite{VB} using a two-component model. The first component is the usual leading
twist
fusion
process  while the second component is based on the model discussed
above.

In the usual leading twist fusion subprocess, there is a
finite probability that the produced charm quark will combine with a
spectator valence quark in the final state to produce a leading hadron.
Such final state coalescence mechanisms have been incorporated into
PYTHIA, a Monte Carlo program based on the Lund string model \cite{Lund}.
In that model, the ``string acceleration" of slow heavy quarks by fast
valence quarks can boost the fast charm rate. However, such a mechanism
 overestimates  the observed asymmetry ${\cal A}(x_F)$ at low $x_F$.
The Lund string model is strictly a final state coalescence. However, the
model we propose is an initial state coalescence. The pion can fluctuate
into higher Fock states as shown in Fig.~14. All the partons have nearly
the same velocity in order to minimize the invariant mass of the state.
As the charm and the valence quark have the same rapidity, it is easy
for them to coalesce to form a large $D$ meson state without paying much
penalty. Thus, it can produce a strong  leading particle correlation
at large $x_F$.

\begin{figure}[htb]
\vspace{2.4in}
\caption {Initial state coalescence  producing a $D$ meson through the
intrinsic charm fluctuation at large $x_{F}$.}
\label{fourteenth-figure}
\end{figure}

Figure 15 shows the results of the two-component model. The parameter $\xi$
determines the relative importance of the leading twist  and
intrinsic charm components. All the calculations
reproduce the general trend of the data.

\begin{figure}[htb]
\vspace{3.0in}
\caption {The prediction of the two-component model. The figure is from
\protect\cite{VB}}
\label{fifteenth-figure}
\end{figure}

\noindent {\rm \bf (e)} \hspace{0.7cm}
The new production mechanisms have all the novel features  observed
in the nuclear dependence of $\jp$ production. QCD factorization is
invalid in the combined limit since there is no relative suppression of
interactions involving several partons in the projectile. The nuclear
$A^{\alpha}$-dependence is a function of $x_F$ rather than a function of
$x_2$ of the target-parton momentum fraction. Because of the rapid
transverse size expansion of the spectators, production cross sections
in nuclear targets becomes surface dominated at large $x_{F}$.

\section{Conclusion}
We have reviewed the experimental status of charm production at large
$x_{F}$ and observed a lot of anomalous behaviors in this kinematic
limit. Both the leading charm asymmetry and the nuclear $\jp$ production show
that factorization breaks down at large $x_{F}$. Higher twist effects
becomes dominant because a new scale $\Lambda^{2}_{QCD}/(1-x_{F})$
emerges, which
reflects the small transverse size of the Fock state, in the
$x \rightarrow 1$ limit. In the combined limit (\ref{new}), the
heavy quark pair can be freed easily as the coherence of the Fock
state is easily broken by soft interactions of  finite transverse
momentum because of the rapid expansion of the transverse size of the
spectators. This new production mechanism helps to explain  the
anomalous phenomena observed at large $x_{F}$.  This new
picture of hadron formation opens up a whole new avenue for studying the
far-off-shell structure of hadrons.  It is thus critical that a new
measurement of the charm and beauty structure functions be performed in
future experiments.


\begin{thebibliography}{99}

\bibitem{BHMT} S. J. Brodsky, P. Hoyer, A. H. Mueller and W.-K. Tang,
\NPB{369}{519}{1992}.

\bibitem{hoyer} P. Hoyer, Acta Phys.Polon. {\bf B23}, 1145 (1992)

\bibitem{EMC} EMC: J. J. Aubert \etal, \NPB{213}{31}{1983}.

\bibitem{CIP} CIP: C. Biino, \etal, \PRL{58}{2523}{1987}.

\bibitem{E537} E537: C. Akerlof, \etal, \PRD{48}{5067}{1993}.

\bibitem{NA3} NA3: J. Badier, \etal, \PLB{114}{457}{1982}.

\bibitem{WA82} WA82: M. Adamovich, \etal, \PLB{305}{402}{1993}.

\bibitem{E769} E769: G. A.Alves, \etal, \PRL{72}{812}{1994}.

\bibitem{Beenakker} W. Beenakker, \etal, \NPB{351}{507}{1991}.

\bibitem{NDE} P. Nason, S. Dawson, and R. K. Ellis, \NPB{327}{49}{1989}.

\bibitem{NA3a} NA3: J. Badier, \etal, Z. Phys. {\bf C20}, 101 (1983).

\bibitem{E772} E772: D. M. Alde, \etal, \PRL{66}{133}{1991}.

\bibitem{E789} E789: M. S. Kowitt, \etal, \PRL{72}{1318}{1994}.

\bibitem{HVS} P. Hoyer, M. V\"anttinen and U. Sukhatme,
\PLB{246}{217}{1990}.

\bibitem{VHBT}  M. V\"anttinen, P. Hoyer, S. J. Brodsky and W.-K. Tang,
in preparation.

\bibitem{VB}  R. Vogt and  S. J. Brodsky, SLAC and LBL preprint,
SLAC-PUB-6468 and LBL-35380, April 1994.

\bibitem{Lund} H.-U. Bengtsson and T. Sj\"{o}strand, Comput. Phys.
Commun. {\bf 46}, 43 (1987).
\end{thebibliography}
\end{document}